\documentstyle[12pt]{article}
\def\p{\partial}
\def\l{\lambda}
\def\bc{\begin{center}}
\def\ec{\end{center}}
\begin{document}
\begin{flushright}
{\sf Preprint JINR E2-94-443, Dubna 1994}\\
hep-th/9508129
\end{flushright}

\vspace{2cm}

\bc{\Large\bf Some new integrable equations from\\
 the self-dual Yang-Mills equations}\ec

\bc {\large T.A.Ivanova\footnote{e-mail: ita@thsun1.jinr.dubna.su} and
A.D.Popov\footnote{e-mail: popov@thsun1.jinr.dubna.su} } \ec
\bc
{\it Bogoliubov Laboratory of Theoretical Physics,\\ JINR, Dubna 141980,
Moscow Region, Russia}\ec

\vspace{1cm}

{\bf Abstract.}

\vspace{0.5cm}

Using the symmetry reductions of the self-dual Yang-Mills (SDYM)
equations in $(2+2)$ dimensions, we introduce new integrable equations
which are nonautonomous versions of the chiral model in $(2+1)$
dimensions, generalized nonlinear Schr\"odinger, Korteweg-de Vries,
Toda lattice, Garnier and Euler-Arnold equations. The Lax pairs for all
of these equations are derived by the symmetry reductions of the Lax
pair for the SDYM equations.

\newpage

{\bf 1. Introduction}

\vspace{0.2cm}

The purpose of this paper is to describe six new systems of differential
equations and to write out the Lax pairs for them. We derive equations
for all these integrable systems using the method of symmetry reduction
(see [1, 2] and references therein) applied to the self-dual Yang-Mills
(SDYM) equations in the space $R^{2,2}$ with the metric of the signature
$(+ + - -)$. For derivation of the Lax pairs for these equations we use the
algorithm of reduction of the Lax pair for the SDYM equations described
in [3].

We use the SDYM equations in $R^{2,2}$ and the symmetry reduction method
only as a tool for obtaining new integrable systems in lower dimensions,
but there are at least three reasons in view of which the connection between
these integrable systems and the SDYM equations is important. Firstly,
the importance of the SDYM equations in $R^{2,2}$  is motivated by the
conjecture [4] that the SDYM equations may be a universal integrable system,
i.e. that all integrable equations in $1\le d\le 3$ dimensions can be obtained
from it by suitable reductions. In fact, it has recently been shown that
many integrable equations can be embedded into the SDYM equations [4--14].
It is obvious that besides the known equations, the symmetry reductions
of the SDYM equations
give  the opportunity to obtain some new integrable equations valuable
for applications [3]. In the following, we illustrate this by deriving
nonautonomous versions of the equations mentioned in the abstract.
 Secondly, to the equations derived from the SDYM equations, one may
apply the twistor techniques for solving equations and for analysing
properties of solutions (see, e.g., [15, 16, 10, 11]).
Thirdly, the SDYM equations are known to arise in the $N=2$ supersymmetric
string theory [17, 18] which is considered now as the universal string theory
including the conventional $N=0$ and $N=1$ strings as particular vacua [19,
20]. Therefore, the soliton-type solutions of the SDYM equations and their
reductions are important for the analysis of nonperturbative effects in
string theories.

\vspace{0.5cm}

{\bf 2. Definitions and notation}

\vspace{0.2cm}

We consider the space $R^{2,2}$ with the metric $(g_{\mu\nu})=
diag(+1, +1, -1, -1)$ and the potentials $A_\mu$ of the Yang-Mills (YM)
fields $F_{\mu\nu}= \p_\mu A_\nu-\p_\nu A_\mu + [A_\mu,A_\nu]$, where $\mu ,
\nu , ... = 1,...,4$, $\p_\mu = \p /\p x^\mu$.  Fields $A_\mu$ and
$F_{\mu\nu}$ take values in the Lie algebra $gl(n, C)$.

In $R^{2,2}$ we introduce null coordinates $t=\frac{1}{2}(x^2-x^4),
u=\frac{1}{2}(x^2+x^4),  y=\frac{1}{2}(x^1-x^3), z=\frac{1}{2}(x^1+x^3)$ and
set $A_{t} = A_{2}-A_{4},\ A_{u} =A_{2}+A_{4},\ A_{y}=A_{1}-A_{3},\
A_{z}=A_{1}+A_{3}$. The SDYM equations in the null coordinates have
the following form:
$$F_{tz}=0,\quad F_{uy}=0,\quad F_{tu}+F_{zy}=0. \eqno(1)$$
Equations (1) can be obtained as compatibility conditions of the
following linear system of equations (cf. ref. [21, 22]):  $$ (\p_t
-\l\p_y + A_t -\l A_y)\Psi =0, \eqno(2a)$$ $$ (\p_z +\l\p_u + A_z +\l
A_u)\Psi =0, \eqno(2b)$$ $$ \p_{\bar\l}\Psi =0, \eqno(2c)$$ where
$\bar\l$ is a complex conjugate to $\l$. Here $\Psi$ is a column
vector depending on the coordinates of $R^{2,2}$ and the
``coordinates" $\l , \bar\l$, parametrizing the upper sheet of the
hyperboloid $H^2=SO(2,1)/SO(2)$. Notice that $\Psi$ is defined on the
twistor space ${\cal Z}=R^{2,2}\times H^2$ for the space $R^{2,2}$,
and eqs.(2) mean the holomorphicity of the vector-function $\Psi$
(Ward theorem [22, 15]).

\vspace{0.5cm}

{\bf 3. Symmetry reduction}

\vspace{0.2cm}

 We consider the inhomogeneous group of rotations $ISO(2,2)$
(rotations and translations)  and an arbitrary subgroup $G$ of the group
$ISO(2,2)$. We would like to impose the conditions of $G$-invariance on the
YM potentials $A_\mu$ and on the vector-function $\Psi$. For that, we have to
define the generators of the group $ISO(2,2)$ as vector fields on $R^{2,2}$
when considering the action of $G$ on $A_\mu$, and as vector fields on the
twistor space $R^{2,2}\times H^2$ when considering the action of $G$ on
$\Psi$ [3].

Let us introduce the following constant tensors:  $$f_{\mu\nu}^{a} =
\{ f_{bc}^{a},\ \mu =a, \nu =b;\ \delta_{\mu}^{a}, \nu =4;\
-\delta_{\nu}^{a}, \mu =4\},\quad I_{a}\,_{\nu}^{\mu} = - \frac{1}{2}
g_{ab} g^{\mu\l}f_{\l\nu}^{b},\eqno(3a)$$ $$\bar f_{\mu\nu}^{a} = \{
f_{bc}^{a},\ \mu =a, \nu =b;\ -\delta_{\mu}^{a}, \nu =4;\
\delta_{\nu}^{a}, \mu =4\},\quad J_{a}\,_{\nu}^{\mu} = - \frac{1}{2}
g_{ab} g^{\mu\l}\bar f_{\l\nu}^{b},\eqno(3b)$$ where $a,b,...=1,2,3,\
g_{11}=g_{22}=-g_{33}=1$ and $f_{23}^{1}=f_{31}^{2}= -f_{12}^{3}=1$
are the structure constants of the group $SO(2,1)$. Then, the
generators of the group $ISO(2,2)$ can be realized in terms of the
following vector fields on $R^{2,2}$:  $$X_a=
I_{a}\,_{\nu}^{\mu}x^\nu\partial_\mu ,\quad Y_a= J_{a}\,_{\nu}^{\mu}
x^\nu\partial_\mu ,\quad P_{\mu}=\p_{\mu}. \eqno(4)$$ The vector
fields on ${\cal Z}=R^{2,2}\times H^2$, which also form the
generators of $ISO(2,2)$, are given by $$\tilde X_a = X_a,\quad
\widetilde{Y}_a= Y_a + Z_a, \quad \widetilde{P}_{\mu}=
P_{\mu},\eqno(5a)$$ with the following expression of the generators
$Z_a$ of the $SO(2,1)$-rotations on $H^2$:
$$Z_1=\frac{1}{2}[(1-\l^2)\p_\l +(1-\bar\l^2)\p_{\bar\l}],\
Z_2=-[\l\p_\l + \bar\l\p_{\bar\l}],\ Z_3=-\frac{1}{2}[(1+\l^2)\p_\l
+(1+\bar\l^2)\p_{\bar\l}]. \eqno(5b)$$ It can be easily shown that
$[X_a,X_b]=f_{ab}^{c} X_c,\ [Z_a, Z_b]= f_{ab}^{c} Z_c,\ [\tilde
Y_a,\tilde Y_b]=f_{ab}^{c} \tilde Y_c$ and so on.

In order to reduce the SDYM equations (1) and the linear system (2)
under a subgroup $G$ of the group $ISO(2,2)$, it is necessary to
impose the following conditions of $G$-invariance on the gauge
potentials $A_\mu$ and on the vector-function $\Psi$ [23]:  $$ W_\xi
A_\mu + A_\sigma W^\sigma_{\xi ,\mu }=0, \quad \forall \xi \in {\cal
G},\eqno(6a)$$ $$ {\tilde W_{\xi}} \Psi =0, \quad \forall \xi \in
{\cal G},\eqno(6b)$$ where $\cal G$ is a Lie algebra of the group
$G$, $W_\xi = W_{\xi}^{\sigma} \p_\sigma$ are vector fields on
$R^{2,2}$ and $\tilde W_\xi =\tilde W_\xi^\sigma\p_\sigma + \tilde
W_\xi^aZ_a$  are vector fields on $R^{2,2}\times H^2$. Both $W_\xi$
and $\tilde W_\xi$ form a realization of the Lie algebra $\cal G$.

In accordance with the general method of symmetry reduction (see [1]
and references therein), as new coordinates on $R^{2,2}\times H^2$, one
should choose the coordinates $\theta_\xi$ on the orbits $Q$ of the group
$G$ in $R^{2,2}\times H^2$, and the
invariant coordinates $\theta_A \ (A=1,..., 4-dim Q)$ and $\zeta$ which
parametrize the space of orbits and satisfy
$$\tilde W_\xi\theta_A=0,\quad
\tilde W_\xi\zeta=0,\quad \p_{\bar\l}\zeta =0,\quad \forall \xi \in
{\cal G}.\eqno(7)$$
Here, the invariant complex coordinate $\zeta$ represents the new ``spectral
parameter". Then,  substituting solutions of eqs.(6) and (7) into eqs.(1),
(2), we obtain the reduced SDYM equations and their Lax pairs in terms of
functions of the invariant coordinates [1, 3].

Now we consider examples of reduction of the SDYM equations to the
integrable equations in $1\le d\le 3$. In what follows, we shall firstly
write out some known integrable equations which arise as reduction of the
SDYM equations under translations. After that we shall describe new
nonautonomous versions of these equations, derived via reduction with
respect to the action of the subgroups containing rotations.

\vspace{0.5cm}

{\bf 4. Reductions to integrable systems in $(2+1)$ dimensions}

\vspace{0.2cm}

{\it Chiral model equation in $R^{2,1}$} [24, 16]. Let us consider the
one-dimensional Abelian group with generator $P_y - P_z$. Then,
$\varphi =y-z$ will be the coordinate on the orbit and the invariant
coordinates are $x=y+z,\ t, \ u$ and $\l$. The YM potentials $A_\mu$,
satisfying (6a), and the
vector-function $\Psi$, satisfying (6b) and (2c), are given by
$$A_t=T_t(t,u,x), A_u=T_u(t,u,x),
A_y=T_y(t,u,x), A_z=T_z(t,u,x), \Psi =\psi(t,u,x,\l ).\eqno(8)$$

Substituting (8) into the linear system (2), we obtain the following
reduced Lax pair:
$$ (\p_t - \l\p_x + T_t - \l T_y)\psi =0,\quad (\p_x + \l\p_u + T_z +
\l T_u)\psi =0.\eqno(9)$$
Accordingly, the SDYM equations (1) are reduced to the
compatibility conditions of the Lax pair (9):
$$ \p_t T_z - \p_x T_t +[T_t, T_z] =0,\quad
\p_x T_u -\p_u T_y + [T_y, T_u]=0, \eqno(10a)$$
$$\p_x (T_y-T_z)+\p_t T_u - \p_u T_t + [T_t, T_u] + [T_z, T_y]=0.
\eqno(10b)$$

Now let us choose the gauge $T_z=T_t=0$. Then, from eqs.(10a) we obtain
$T_u=g^{-1}\p_u g, \ T_y=g^{-1}\p_x g$, where $g$ is an arbitrary
function of $t, u, x$ with values in the group $GL(n, C)$, and eqs. (10b)
coincide with the equations of the chiral field model considered in
the papers [24, 16]:
$$\p_x(g^{-1}\p_x g) + \p_t(g^{-1}\p_u g) =0. \eqno(11)$$

{\it Nonautonomous chiral model equation in $R^{2,1}$}. Now we consider the
one-dimensional Abelian group of rotations generated by the vector field
$X_2+Y_2$. From (4) and (5), we obtain $\tilde X_2+\tilde Y_2 = X_2+Y_2+Z_2=
z\p_z - y\p_y -\l\p_\l-\bar\l\p_{\bar\l}$. Let us
introduce  the coordinates $\rho , \theta , \eta , \xi $ by formulae
$y=\frac{1}{2}\rho\,e^{-\theta},\ z=\frac{1}{2}\rho\,e^{\theta},\ \l
=\eta\,e^{i\xi}$, then $X_2+Y_2=\p_\theta$ and $\tilde X_2 +\tilde Y_2
=\p_\theta -\eta\p_\eta$.
Therefore, $\varphi = \frac{1}{2}(\theta - \ln\eta)$ will be the
coordinate on the orbit and $t, u, \rho , \zeta =\l e^\theta$
will be the invariant coordinates.

The invariant YM potentials $A_\mu$, satisfying eqs.(6a), have the form
$$A_t=T_t(t,u,\rho ),\  A_u=T_u(t,u,\rho ),\  A_y=T_y(t,u,\rho )e^\theta ,\
A_z=T_z(t,u,\rho)e^{-\theta }.\eqno(12a)$$
The vector-function
$$\Psi =\psi (t, u,\rho , \zeta )\eqno(12b)$$ is the solution of equations
(6b) and (2c).

Substituting (12) into (2), we obtain the following reduced Lax pair:
$$\nabla_{V_1}\psi\equiv [\p_t - \zeta\p_\rho + \frac{1}{\rho
}\zeta^2\p_\zeta + T_t - \zeta T_y]\psi = 0,\ \nabla_{V_2}\psi
\equiv [\p_\rho
+\zeta \p_u + \frac{1}{\rho }\zeta\p_\zeta + T_z + \zeta T_u]\psi =
0,\eqno(13)$$
where $V_1=\p_t - \zeta\p_\rho +\frac{1}{\rho }\zeta^2\p_\zeta
,\ V_2=\p_\rho +\zeta \p_u + \frac{1}{\rho }\zeta\p_\zeta$.
Remind that in the general case $[V_1, V_2]\ne 0$ and then for
linear systems like (13) the compatibility condition is
$$[\nabla_{V_1}, \nabla_{V_2}]-\nabla_{[V_1, V_2]}=0.\eqno(14)$$
Correspondingly, the SDYM equations (1) are reduced to
$$ \p_t T_z - \p_\rho T_t + [T_t, T_z]
=0,\quad \p_\rho T_u -\p_u T_y + [T_y, T_u]=0, \eqno(15a)$$
$$\p_\rho (T_y-T_z)+\frac{1}{\rho}(T_y-T_z) + \p_t T_u - \p_u T_t +
[T_t, T_u] + [T_z,T_y]=0 \eqno(15b)$$
which agree with the compatibility condition (14) of the Lax pair (13).

Choosing the same gauge $T_z=T_t=0$ as in (9), (10), from
eqs.(15a) we obtain $T_u=g^{-1}\p_u g, T_y=g^{-1}\p_\rho g$. Then, eq.(15b)
is reduced to the nonautonomous chiral model equation in $R^{2,1}$:
$$\p_\rho
(g^{-1}\p_\rho  g)+\frac{1}{\rho}g^{-1}\p_\rho g+\p_t(g^{-1}\p_u g) =0
\Longleftrightarrow  \frac{1}{\rho}\p_\rho (\rho g^{-1}\p_\rho  g) +
\p_t(g^{-1}\p_u g) =0. \eqno(16)$$
The Lax pair for this equation has the form (13) with $T_z=T_t=0$.

{\it Remark}. Notice that if one uses an additional condition of
invariance under $P_t+P_u:\ (\p_t+\p_u)\psi =(\p_t+\p_u)g=0$ in the
Lax pair (9) and in eqs.(11), then one obtains the equation of the
principal chiral model in $R^{1,1}$. But if we impose the same
condition on $\psi$ and $g$ in (13) and (16), then we obtain the
nonautonomous equation of the principal chiral model in $R^{1,1}$
[25, 26], which in a particular case of the gauge group $GL(2,R)$ is
equivalent to the Ernst equations [25, 10].

\vspace{0.5cm}

{\bf 5. Reductions to integrable systems in $(1+1)$ dimensions}

\vspace{0.2cm}

{\it Generalized nonlinear Schr\"odinger equation (NLS)} [9].
Let us consider the two-dimensional Abelian group with the generators
$\{P_y-P_z, P_u\}$.  Then, solutions $A_\mu$ and $\Psi$ of eqs.(6) and
(2c) are given by
$$A_t = T_t (t, x),\ A_u=T_u(t,x),\ A_y=T_y(t, x),\ A_z=T_z(t, x),\
\Psi=\psi (t,x,\l ).\eqno(17)$$
The linear system (2) is reduced to the following one:
$$\left\{{(\p_t-\l \p_x + T_t -\l  T_y)\psi =0, \atop (\p_x+T_z+\l T_u)\psi
=0}\right.\ \Rightarrow \left\{ {[\p_t + T_t + \l  (T_z-T_y) +\l ^2T_u]\psi
=0,\atop (\p_x+T_z+\l T_u)\psi =0}\right.\eqno(18)$$
and the SDYM equations (1) are reduced to the compatibility condition of the
Lax pair (18):
$$\p_{t}T_{z}-\p_{x}T_{t}+[T_{t},T_{z}]=0,\quad \p_{x}T_{u}-[T_{u},T_{y}]=0,
\eqno(19a)$$
$$\p_{t}T_{u}-\p_{x}(T_{z}-T_{y})+[T_{t},T_{u}]+[T_{z},T_{y}]=0.\eqno(19b)$$

To reduce eqs.(19) to the generalized NLS equations, introduced by Fordy and
Kulish [27], one should impose the algebraic constraints on the elements of
matrices in (19). Let us choose in $GL(n, C)$ the subgroups $N$ and $H$
so that $N/H$ be a compact Hermitian symmetric space.
Let ${\cal N}$ and ${\cal H}$ be the Lie algebras of the
Lie groups $N$ and $H$. Then ${\cal N} = {\cal H} \oplus  {\cal P}$
and $[{\cal H},{\cal H}]\subset {\cal H}, [{\cal H},{\cal P}]\subset
 {\cal P}, [{\cal P},{\cal P}]\subset {\cal H}$.  A special feature
 of Hermitian symmetric spaces is the existence  of an element $A\in
{\cal H}$ such that ${\cal H} = \{ B\in {\cal N}: [A,B]=0\}$. The matrix
ad$_{A}$  has only three distinct eigenvalues $0, \pm i$  and
$[A,{\cal H}]=0, [A,X^{\pm }]= \pm iX^{\pm }$ for all $X^{\pm }\in
{\cal P}^{\pm }, {\cal P}={\cal P}^{+}\oplus  {\cal P}^{-}$.
Let $e_{\pm \alpha }$ be a basis of the space ${\cal P}^{\pm }$. Then
$$
[A, e_{\pm \alpha }]=\pm i e_{\pm \alpha },\quad
[e_\mu , [e_\nu , e_{-\sigma }]]= R_{\mu ,\nu ,-\sigma }^\alpha e_\alpha
$$
$$
[e_{-\mu} , [e_{-\nu} , e_{\sigma }]]= R_{-\mu ,-\nu ,\sigma }^{-\alpha}
e_{-\alpha}, \quad
R_{-\mu ,-\nu ,\sigma }^{-\alpha} =\bar R_{\mu ,\nu ,-\sigma }^\alpha ,
\eqno(20)
$$
where $R_{\mu ,\nu ,-\sigma }^\alpha $ are components of the curvature
tensor defined at the initial point of the
symmetric space $N/H$, and $R_{-\mu ,-\nu ,\sigma }^{-\alpha}$ are
complex conjugate to the $R_{\mu ,\nu ,-\sigma }^\alpha $ components.

For the matrices from (17) we choose the following ansatz:
$$
T_t=\sum_\alpha (\phi^\alpha e_\alpha + \bar \phi^\alpha e_{-\alpha }) +
\sum_{\alpha ,\beta}\Omega^{\alpha , -\beta}[e_\alpha , e_{-\beta }],\quad
T_u=A,$$
$$ T_y=0,\quad T_z=\sum_\alpha(\psi^\alpha e_\alpha +
\bar\psi^\alpha e_{-\alpha }), \eqno(21)$$
where $\phi^\alpha , \psi^\alpha$ and $\Omega^{\alpha , -\beta}$ are
arbitrary complex-valued functions of $t, x$ and the bar over the letter
means complex conjugation. Substituting (21) into eqs.(19), we obtain that
$$
\phi^\alpha = i\p_x\psi^\alpha ,\ \Omega^{\alpha , -\beta} = i(\psi^\alpha
\bar\psi^\beta + \Omega^{\alpha , -\beta}_o), \
\Omega^{\beta , -\alpha}_o = \bar \Omega^{\alpha , -\beta}=const
\eqno(22)$$
and eqs.(19) are reduced to the generalized NLS equations on the functions
$\psi^\alpha$:
$$
i\p_t\psi^\alpha + \p^2_x\psi^\alpha +
\sum_{\mu ,\nu ,\sigma }R_{\mu ,\nu ,-\sigma }^\alpha \psi^\mu
\psi^\nu\bar\psi^\sigma  +
\sum_{\mu ,\nu ,\sigma }R_{\mu ,\nu ,-\sigma }^\alpha
\Omega_o^{\nu ,-\sigma }\psi^\mu =0.
\eqno(23)$$
Notice that the constant components $\Omega_o^{\nu ,-\sigma }$ can always
be chosen so that  $\sum_{\nu ,\sigma }R_{\mu ,\nu ,-\sigma }^\alpha
\Omega_o^{\nu ,-\sigma }=\omega_\alpha\delta^\alpha_\mu $,
where $\omega_\alpha$ are real constants [27].
The Lax pair for eqs.(23) can be deduced via substitution of (21) and
(22) in (18).

{\it Nonautonomous generalized NLS equation}. Now let us consider the
two-di\-men\-sio\-nal Abe\-li\-an group with the generators
$\{X_2+Y_2, P_u\}$. Then, invariant $A_\mu$ and $\Psi$ are
given by formulae (12) where  $T_\mu$ and $\psi$ do not depend on $u$.
The reduced Lax pair and SDYM equations have the form
$$[\p_t -\zeta\p_\rho +\frac{1}{\rho }\zeta^2\p_\zeta + T_t - \zeta T_y]
\psi = 0,\quad [\p_\rho +\frac{1}{\rho}\zeta\p_\zeta +T_z+\zeta T_u]
\psi =0,\eqno(24)$$
$$
\p_t T_z-\p_\rho T_t+[T_t,T_z]=0,\quad \p_\rho T_u+[T_y,T_u]=0,\eqno(25a)$$
$$\p_\rho (T_y-T_z)+\frac{1}{\rho}(T_y-T_z)+ \p_t T_u
+[T_t, T_u]+[T_z, T_y]=0. \eqno(25b)$$

For matrices from (24), (25) we choose the ansatz (21) again.
Substituting (21) into (25), we obtain that
$$
\phi^\alpha = i(\p_\rho\psi^\alpha + \frac{1}{\rho}\psi^\alpha ), $$
$$\Omega^{\alpha , -\beta} = i(\psi^\alpha\bar\psi^\beta +
\Omega^{\alpha , -\beta}_o +2\int \frac{d\rho}{\rho}\psi^\alpha\bar
\psi^\beta ),
\ \Omega^{\beta , -\alpha}_o=\bar \Omega^{\alpha , -\beta}_o=const
\eqno(26)$$
and the functions $\psi^\alpha$ have to satisfy the nonautonomous
generalized NLS equations
$$
i\p_t\psi^\alpha + \p^2_\rho\psi^\alpha +
\sum_{\mu ,\nu ,\sigma }R_{\mu ,\nu ,-\sigma }^\alpha \psi^\mu
\psi^\nu\bar\psi^\sigma  + \sum_{\mu ,\nu ,\sigma }
R_{\mu ,\nu ,-\sigma }^\alpha \Omega_o^{\nu ,-\sigma }\psi^\mu = $$
$$
= -\p_\rho(\frac{1}{\rho}\psi^\alpha ) - 2
\sum_{\mu ,\nu ,\sigma }R_{\mu ,\nu ,-\sigma }^\alpha \psi^\mu
\int\frac{d\rho}{\rho}\psi^\nu\bar\psi^\sigma  .\eqno(27)$$
The Lax pair for eqs.(27) can be obtained by substitution of (21)
and (26) into (24).

{\it Remark}. In the case of $N=SU(2)$ and $H=U(1)$, ansatz (21) has
the form
$$T_t=\pmatrix{\Omega  &\bar \phi  \cr \phi  &-\Omega  },\
T_u=\frac{1}{2i}\pmatrix{1&0\cr 0&-1},\
T_y =0,\ T_z=\sqrt{\kappa}\pmatrix{0&\bar \psi \cr \psi &0},\eqno(28)$$
where $\Omega , \phi$ and $\psi$ are arbitrary complex-valued functions
of $t$ and $\rho$,
and $\kappa$ is an arbitrary real constant parameter. Substituting (28)
into (25), we obtain that
$$
\Omega = -i \kappa (\bar\psi \psi -\gamma^2) -
2i\kappa \int \frac{d\rho}{\rho}\bar\psi \psi ,
\qquad \phi = i\sqrt{\kappa}(\p_\rho \psi +\frac{1}{\rho}\psi ),\quad
\gamma =const, \eqno(29)$$
and the function $\psi$ has to satisfy the equation
$$i\p_t\psi +\p^2_\rho\psi - 2\kappa
(\bar\psi\psi -\gamma^2)\psi = -\p_\rho (\frac{1}{\rho} \psi )+
 4\kappa\psi \int\frac{d\rho}{\rho}\bar\psi\psi .
\eqno(30)$$
The Lax pair for eqs.(30) can be obtained by substitution of (28) and (29)
into (24).

The nonautonomous NLS equation (30) has been considered in the paper
[26]. When $\kappa =-1$ and $\gamma^2=0$, this equation is gauge equivalent
to the equation of
the Heisenberg ferromagnet in axial geometry. By change of variables $t,
\rho$
and $\psi$, eq.(30) can be transformed to the equation, which has been
introduced and integrated in [28]. Thus, the nonautonomous NLS equation
is shown to be the reduction of the SDYM equations.

{\it Korteweg-de Vries equation} [5,6]. Now, considering the
generators $\{P_y-P_z, P_u\}$, the Lax pair (18) and the compatibility
conditions (19), we choose
the matrices from (19) in the form of the following $2\times 2$ matrices
$$T_t=\pmatrix{a&b\cr c&-a},\ T_u=\pmatrix{0&0\cr -1&0},\
T_y =\pmatrix{0&0\cr h&0},\ T_z=\pmatrix{0&g\cr f&0},\eqno(31a)$$
where $a, b, c, f, g$ and $h$ are arbitrary real-valued functions.

Substituting (31a) in (19), we obtain that
$$a=\frac{1}{4}\p_xf,\ b=-\frac{1}{2}f,\ c=\frac{1}{2}f^2 +
\frac{1}{4}\p_{x}^{2}f, \ h=\frac{1}{2}f,\ g=-1,\eqno(31b)$$
and the function $f$ has to satisfy the KdV equation
$$\p_tf-\frac{3}{2}f\p_xf - \frac{1}{4}\p_{x}^{3}f=0.\eqno(32)$$
The Lax pair for eqs.(32) is obtained after substitution of (31a) and
(31b) into (18).

{\it Nonautonomous KdV equation}. Now we consider the generators
$\{X_2+Y_2, P_u\}$, Lax pair (24) and its compatibility conditions (25).
For matrices from (25) let us choose the ansatz (31a). Substituting (31a)
in (25), we obtain that
$$a=\frac{1}{4}\p_\rho f -\frac{1}{4\rho}\int\frac{d\rho}{\rho}f,\quad
b=-\frac{1}{2\rho}f -\frac{1}{2\rho}\int\frac{d\rho}{\rho}f,\quad
h=\frac{1}{2}f + \frac{1}{2}\int\frac{d\rho}{\rho}f,$$ $$c=\frac{1}{4}\rho
\p_{\rho}^{2}f -\frac{1}{4\rho}f +\frac{1}{2}f^2+(\frac{1}{4\rho}+\frac{f}
{2})\int\frac{d\rho}{\rho}f,\quad g=-\frac{1}{\rho},\eqno(33)$$
and the function $f$ satisfies the equation
$$\p_tf-\frac{3}{2}f\p_\rho f - \frac{1}{4}\p_{\rho}^{3}(\rho f) =$$
$$=\frac{1}{2\rho^2}f + \frac{1}{2\rho}f^2 - \frac{1}{4\rho}\p_\rho f -
\frac{1}{2}\p_{\rho}^{2}f + (\frac{1}{2}\p_\rho f - \frac{f}{2\rho}-
\frac{1}{4\rho^2})\int\frac{d\rho}{\rho}f.\eqno(34)$$
The Lax pair for eq.(34) is obtained after substitution of (31a) and
(33) into the Lax pair (24).

{\it Remark.} Nonautonomous KdV equations have been considered in the
papers [26, 28]. Equation (34) differs from ones, considered in [26,
28], and it is a new deformation of the KdV equation.

\vspace{0.5cm}

{\bf 6. Reductions to integrable dynamical systems}

\vspace{0.2cm}

{\it Periodic Toda lattice with damping}. Let us consider the
three-dimensional non-Abelian subgroups of $ISO(2,2)$ generated by
the vector fields $X_2 + \beta Y_2, P_y, P_z$, where
$\beta \in R, \beta\ne 1$. Notice that the SDYM equations,
reduced with respect
to the symmetry group with the generators $X_2+Y_2, P_y$ and $P_z$, lead
to the zero curvature condition $F_{\mu\nu}=0$ and, therefore, they are not
interesting. That is why we shall investigate the case $\beta\ne 1$.

Let us introduce the coordinates $\tau , \theta$ by formulae
$\tau =\frac{1}{4}\ln (4tu)^2, \ \theta =\frac{1}{4}\ln (\frac{u}{t})^2$.
Then, the orbit coordinates are $\chi =\frac{2(1+\beta^2)}{(1 -\beta )}
\theta +\frac{1}{2}\beta \ln (\bar\l\l ),\ y,\ z\ $ and the invariant
coordinates are $\tau ,\ \zeta =\l e^{\gamma\theta },$ where $\gamma =
2\beta/(1 -\beta )$. The invariant YM potentials and $\Psi$ satisfying
eqs.(2c) and (6) are given by
$$A_t=T_t(\tau )e^{\theta -\tau },\  A_u=T_u(\tau )e^{-\theta -\tau },\
A_y=T_y(\tau )e^{(1+\gamma )(\theta -\tau )},\
A_z=T_z(\tau )e^{-(1+\gamma )(\theta +\tau )}$$
$$\Psi =\psi (\tau , \zeta ). \eqno(35)$$

Substituting (35) into the linear system (2), changing the variables and
using (6), we obtain the following reduced Lax pair:
$$
[\p_\tau - \gamma\zeta\p_\zeta +T_t - \zeta e^{-\gamma\tau}T_y]\psi =0,
\quad
[\zeta\p_\tau + \gamma\zeta^2\p_\zeta + e^{-\gamma\tau}T_z +\zeta T_u]\psi
=0, \eqno(36)$$
Using the compatibility condition (14) for the Lax pair (36), we obtain
the following reduced SDYM equations:
$$
\frac{d}{d\tau}T_y + [T_u, T_y]=0,\quad  \frac{d}{d\tau}T_z + [T_t, T_z]=0,
\eqno(37a)$$
$$\frac{d}{d\tau}\left (T_u-T_t\right )+[T_t,T_u]+ e^{-2\gamma\tau}
[T_z,T_y]=0. \eqno(37b)$$

The equations of the periodic Toda lattice with damping are derived via
the algebraic reduction of eqs.(37).
Let us choose for $T_t, T_u, T_y, T_z \in gl(n, C)$ the following
(algebraic)  ansatz:
$$
T_t= -T_u = \pmatrix{p_1&0&...&0\cr 0&p_2&\ddots &\vdots\cr
\vdots&\ddots&\ddots&0\cr 0&\ldots&0&p_n},\quad
T_y= T_z^T =2 \pmatrix{0&a_1&0&...&0\cr 0&0&a_2&\ddots &\vdots\cr
\vdots&\ddots&\ddots&\ddots&0\cr 0&0&\ldots&0&a_{n-1}\cr a_n&0&\ldots&0&0},
\eqno(38)$$
where $a_\alpha =\exp(q_\alpha -q_{\alpha +1})$ and the superscript $^T$
means
matrix transpose. Then, after substitution of (38) into eqs.(37), we obtain
$$\frac{d}{d\tau}q_\alpha =p_\alpha ,\quad \frac{d}{d\tau}p_\alpha =
2\exp(-2\gamma\tau )\left\{ \exp [2(q_{\alpha -1}-q_\alpha )] -
\exp [2(q_{\alpha }-q_{\alpha +1})]\right\}.
\eqno(39)$$
When $\gamma =0$, the latter equations coincide with the standard
periodic Toda lattice equations. If $\gamma\ne 0$, then using the variable
$\varphi =\exp (-\gamma\tau )$, we obtain the equations of the Toda lattice
with damping:
$$\frac{d^2}{d\varphi ^2}q_\alpha +\frac{1}{\varphi }
\frac{d}{d\varphi }q_\alpha =
\frac{2}{\gamma^2}\left\{ \exp [2(q_{\alpha -1}-q_\alpha )] -
\exp [2(q_{\alpha }-q_{\alpha +1})]\right\}.
\eqno(40)$$
The corresponding Lax pair is obtained by inserting (38) in (36).

{\it Integrable Hamiltonian systems with quartic potentials}. We are still
considering the generators $X_2 +\beta Y_2, P_y, P_z$, the Lax pair
(36) and the compatibility conditions (37). Let us choose in $GL(n, C)$
the subgroups $N$ and $H$ in such a way that $N/H$ be the Hermitian
symmetric space (for definitions and notation see Sec.5).

For the matrices from (37) we choose the following ansatz:
$$T_{t} = 0,\quad T_u=i\sum_{\alpha}q^{\alpha}(e_{\alpha}+e_{-\alpha}),$$
$$T_{y}=\sum_{\alpha } r^{\alpha }(e_{\alpha }-e_{-\alpha })+
\sum_{\alpha ,\sigma } \Omega^{\alpha ,-\sigma}[e_{\alpha },e_{-\sigma }],
\quad
T_{z} = A,\eqno(41) $$
where $q^{\alpha }$, $r^{\alpha }$ and $\Omega^{\alpha ,-\sigma}$ are
arbitrary real-valued functions  of $\tau$.

Substituting (41) in (37) we obtain that
$$r^\alpha = -e^{2\gamma\tau}\frac{dq^\alpha}{d\tau},\ \Omega^{\alpha ,
-\sigma} = i\Omega_o^{\alpha ,-\sigma} -i e^{2\gamma\tau}q^\alpha q^\sigma
+2i\,\gamma\int d\tau\ e^{2\gamma\tau} q^\alpha q^\sigma ,\
\Omega_o^{\alpha ,-\sigma}=const, \eqno(42)$$
and eqs.(37) are reduced to the equations
$$
\frac{d^2}{d\tau^2} q^{\alpha } -
\sum_{\mu ,\nu ,\sigma  } R^{\alpha}_{\mu ,\nu ,- \sigma   }
q^{\mu }q^{\nu }q^{\sigma  }+
\sum_{\mu ,\nu ,\sigma  } R^{\alpha}_{\mu ,\nu ,-\sigma   }
\Omega_o^{\nu ,-\sigma } q^{\mu }=$$
$$=-2\gamma \left (\frac{dq^\alpha}{d\tau}+e^{-2\gamma\tau}
\sum_{\mu ,\nu ,\sigma  } R^{\alpha}_{\mu ,\nu ,-\sigma }q^{\mu }
\int d\tau\ e^{2\gamma\tau}q^{\nu }q^{\sigma  }\right )+
(1-e^{-2\gamma\tau})
\sum_{\mu ,\nu ,\sigma  } R^{\alpha}_{\mu ,\nu ,-\sigma   }
\Omega_o^{\nu ,-\sigma } q^{\mu }.
\eqno(43)$$
Notice that $\Omega_{o}^{\alpha , -\sigma}$ may always be chosen so that
$\sum_{\nu ,\sigma  } R^{\alpha}_{\mu ,\nu ,-\sigma}\Omega_o^{\nu ,
-\sigma } = \omega_\mu\delta_{\mu}^{\alpha}$, where $\omega_\mu
=const$.

When $\gamma =0$, eqs.(43) coincide with the equations of motion in
quartic potentials, considered in [29]. Equations of the Garnier
system are the particular case of eqs.(43), corresponding to
$\gamma =0, N=SU(n), H=S(U(1)\times U(n-1)).$ The Lax pair for
eqs.(43) can be obtained by inserting (41) and (42) into (36).

{\it Euler-Arnold equations and their deformations.} Now let us consider
the three-dimen\-sio\-nal non-Abelian symmetry group with the generators
$\alpha X_2 +\beta Y_2,\ P_u, \ P_y$, where  $\alpha ,\beta\in R, \
\alpha^2-\beta^2=1$. Let us introduce the coordinates $\tau =\frac{1}{2}
(\alpha -\beta )\ln z^2+ \frac{1}{2}(\alpha +\beta )\ln t^2,\ \theta =
\frac{1}{2}(\alpha -\beta )\ln z^2 -\frac{1}{2}(\alpha +\beta )\ln t^2$.
In this case the orbits are parametrized by the coordinates  $\chi =
\theta -\frac{1}{2}\beta\ln (\bar\l\l ),\ u,\ y\ $ and the invariant
coordinates are $\tau ,\ \zeta = \l e^{\beta\theta}$.
Solving eqs. (6) and (2c), we obtain the following formulae for the
invariant YM potentials and for the vector-function $\Psi$
$$A_t=(\alpha +\beta )\exp \left (\frac{\theta -\tau}{2(\alpha +\beta )}
\right )T_t(\tau ),\
A_u=(\alpha -\beta )\exp \left (-\frac{\tau}{2(\alpha -\beta )}
-\frac{\theta}{2(\alpha +\beta )}\right )T_u(\tau ),$$
$$
A_y=(\alpha +\beta )\exp \left (-\frac{\tau}{2(\alpha +\beta )}
+\frac{\theta}{2(\alpha -\beta )}\right )T_y(\tau ),\
A_z=(\alpha -\beta )\exp \left (\frac{-\theta -\tau}{2(\alpha -\beta )}
\right )T_z(\tau ),$$
$$\Psi =\psi (\tau , \zeta ).\eqno(44)$$

Substitute (44) into the linear system (2) and express the
derivatives in (2) via the new coordinates. Then, after using the
conditions of invariance of $\psi$, the linear system (2) is
reduced to the following one:
$$(\p_\tau -\beta\zeta\p_\zeta +T_t - \zeta T_y )\psi =0,\quad
 (\p_\tau + \beta\zeta\p_\zeta + T_z + \zeta T_u )\psi =0.\eqno(45)$$
If we put $N_t=\frac{1}{2}(T_z+T_t),\ N_u=\frac{1}{2}(T_y+T_u),\
N_y=\frac{1}{2}(T_y-T_u),\ N_z=\frac{1}{2}(T_z-T_t)$,
then we can rewrite the Lax pair (45) in the form
$$(\p_\tau +N_t  - \zeta N_y )\psi =0, \quad (\beta\zeta\p_\zeta +N_z +
\zeta N_u)\psi =0.\eqno(46)$$
The compatibility conditions of the Lax pair (46) are
$$\p_\tau N_u+\beta N_y+[N_t,N_u]+[N_z,N_y]=0,\quad [N_u,N_y]=0
\eqno(47a)$$
$$\p_\tau N_z+[N_t,N_z]=0.\eqno(47b)$$

Let us choose
$N_t$, $N_z$ to be antisymmetric $n\times n$ matrices and
$N_u$, $N_y$ to be diagonal matrices satisfying the
equation $\p_\tau N_u+\beta N_y=0$.
We choose the solution of this equation in the
following form: $N_u={\cal A} - \beta \tau N_y$,
where ${\cal A}=diag(a_1,...,a_n)$ and $N_y= diag(b_1,...,b_n)$ are constant
diagonal matrices with $a_i\not=a_j$, $b_i\not=b_j$ and $a_i\not=b_j$
when $i\ne j$. Solution of eqs.(47a) may be written in the form
$$N_u={\cal A}-\beta \tau N_y,
 \quad (N_t)_{ij}=\l_{ij}M_{ij},\quad
(N_z)_{ij}=(\beta\tau\l_{ij}-1)M_{ij},\eqno(48)$$
where $\l_{ij}=(b_i-b_j)/(a_i-a_j)$, $M=(M_{ij})=(-M_{ji})$ is an
arbitrary antisymmetric matrix and $(N_t)_{ij},\ (N_z)_{ij}$ are
components of the matrices $N_t, N_z$.

After substitution of (48), eqs.(47b) can be written in the form
$$
\frac{d}{d\tau}[(1- \beta\tau\l_{ij})M_{ij}]=\sum_k(\l_{kj}-\l_{ik})
M_{ik} M_{kj}.\eqno(49)$$
When $\beta = 0$, these equations coincide with the
Euler-Arnold equations describing the rotation of the $n$-dimensional
rigid body [30]. One may also rewrite eqs.(49) in the form
$$\frac{d}{d\tau}W_{ij}=\sum_k(\Lambda _{kj}-\Lambda_{ik})W_{ik}W_{kj},
\eqno(50)$$
where $W_{ij}=(1-\beta\tau\l_{ij})M_{ij}$ and $\Lambda_{ij}(\tau )=
\l_{ij}/(1-\beta \tau\l_{ij})$.

{\it Generalized Calogero-Moser system}.
Here we consider the same symmetry group with the generators
$\alpha X_2+\beta Y_2,\ P_u, \ P_y\ ( \alpha ,\beta\in R,\
\alpha^2-\beta^2=1)$,
the Lax pair (46) and the compatibility conditions (47).
Now for the matrices $N_t, N_u, N_y, N_z \in gl(n, C)$ we choose the
following algebraic ansatz:
$$N_t = -i\sum_{j,k=1\atop j\ne k}^{n}\frac{f_{j}^{+}f_k}{(q_j-q_k)^2}
e_{jk}-2N_z,\
\quad N_u = -i\sum_{j,k=1}^{n}f_{j}^{+}f_k e_{jk}, \quad N_y=2N_u,$$
$$
N_z = h\left\{\sum_{k=1}^{n}p_k e_{kk} + i\sum_{j,k=1\atop j\ne k}^{n}
\frac{f_{j}^{+}f_k}{q_j-q_k}e_{jk}\right\}.\eqno(51)$$
Here $p_j, q_j$ and $h$ are the real-valued functions of $\tau$,
the vector-functions $f_j(\tau )$ belong to $N$-dimensional vector
space $C^N$, $f^+_j$ are canonical
conjugate to $f_j$: $f^+_jf_j=\sum_{a=1}^{N}\bar f^a_j f^a_j =1$, and
$(e_{jk})_{mn}=\delta_{jm}\delta_{kn}$ are the generators of the group
$SL(n, C)$: $[e_{jk},e_{lm}]=\delta_{kl}e_{jm}-\delta_{mj}e_{lk}$.

Substituting (51) into (47), we obtain that $h=\exp (2\beta\tau )$ and
eqs.(47) are reduced to the following system of equations:
$$
\frac{d}{d\tau}q_j=p_j,\
\frac{d}{d\tau}p_j + 2\beta p_j =
2\sum_{k=1\atop k\ne j}^{n}\frac{(f^+_jf_k)(f^+_kf_j)}{(q_j-q_k)^3},$$
$$
\frac{d}{d\tau}f_j + \beta f_j = -i\sum_{k=1\atop k\ne j}^{n}
\frac{f_k(f^+_kf_j)}{(q_j-q_k)^2},\quad
\frac{d}{d\tau}f_j^+ + \beta f_j^+ = i\sum_{k=1\atop k\ne j}^{n}
\frac{(f^+_jf_k)f^+_k}{(q_j-q_k)^2} \ . \eqno(52)$$
The Lax pair for eqs.(52) can be obtained by substituting (51) into (46).
When $\beta =0$, eqs.(52) coincide with those of the generalized
Calogero-Moser system introduced in [31] (see also [32] and references
therein). It may be shown that by change of variables $\tau , p_j$ and
 $f_j$, eqs.(52) can be transformed to the standard equations with
$\beta =0$.

{\it Euler-Calogero-Moser system.} Considering the same symmetry
group with the generators $\alpha X_2+\beta Y_2,\ P_u, \ P_y \quad
(\alpha ,\beta\in R,\ \alpha^2-\beta^2=1)$, the Lax pair (46) and its
compatibility conditions (47), for matrices in (46), (47) we choose
now the following ansatz:
$$
N_t=\sum_{j,k=1\atop j\ne k}^{n}\frac{h_{jk}}{(q_j-q_k)^2}\,e_{jk} - 2N_z,
\quad
N_u=-\sum_{j,k=1}^{n}h_{jk}e_{jk},\quad N_y=2N_u,$$
$$N_z=\exp (2\beta\tau )\left\{\sum_k p_ke_{kk} -
\sum_{j,k=1\atop j\ne k}^{n}\frac{h_{jk}}{q_j-q_k}\,e_{jk}\right\},
\eqno(53)$$
where $p_j, q_j$ and $h_{ij}=-h_{ji}$ are the real-valued functions
of $\tau $, and $e_{jk}$ are the generators of the group $SL(n, R)$.

Substituting (53) into eqs.(47) yields
$$
\frac{d}{d\tau}q_j=p_j,\
\frac{d}{d\tau}p_j + 2\beta p_j =
2\sum_{k=1\atop k\ne j}^{n}\frac{h_{jk}h_{jk}}{(q_j-q_k)^3},$$
$$
\frac{d}{d\tau}h_{jk} + 2\beta h_{jk} =
\sum_{m=1\atop m\ne j,k}^{n}h_{jm}h_{mk}
\left [\frac{1}{(q_k-q_m)^2}-\frac{1}{(q_j-q_m)^2}\right ].
\eqno(54)$$
The Lax pair for eqs.(54) has the form (46) with the matrices
$N_t, N_u, N_y, N_z$ from (53). When $\beta =0$, eqs.(54) coincide with
those of the Euler-Calogero-Moser system introduced in the paper [33]
(see also the discussion of this integrable system in [32]).
It may be shown that by change of variables $\tau , p_j$ and $h_{ij}$,
eqs. (54) can be transformed to the standard equations with $\beta =0$.

\vspace{0.5cm}

{\bf 7. Conclusion}

\vspace{0.2cm}

In this paper, we have introduced six new systems of nonlinear
integrable differential
equations. The Lax pairs for all of these systems contain derivatives
of the form $\zeta\frac{\p}{\p\zeta}$ with respect to the spectral
parameter
$\zeta$. The differential operator $\zeta\frac{\p}{\p\zeta}$ corresponds
to the extention of the loop algebra associated with the Lie algebra of
the gauge group, and this extention is important in the standard approach
to the integrable equations in $(1+1)$ dimensions [34].
 The dressing method for the Lax pairs containing the additional term
$\zeta\p_\zeta$
has been developed by Belinsky and Zakharov [25]. Namely, if one chooses a
``seed solution" of an integrable system and constructs
the corresponding solution $\psi_0$ of the linear system, then the ansatz
for iteration  is $\psi_n=(I+\frac{R_n}{\zeta -\mu_n})\psi_{n-1}$,
where the matrices $R_n$ are independent of $\zeta$, and $\mu_n$ are
functions of the coordinates and do not depend on $\zeta$ ({\it moving}
poles). For more detailed discussions see, e.g., [25, 26, 34].

\vspace{0.5cm}

{\bf Acknowledgements}

One of us (A.D.P.) thanks for support and hospitality the Max-Planck-Institut
f\"{u}r Physik, M\"{u}nchen, where part of this work was done.
This work was supported by the International Science Foundation
(grant \# NK8300), by the Russian Foundation for Fundamental Research
(grant \# 95-01-00027) and by the Heisenberg-Landau Program.

\newpage

{\bf References}
\begin{enumerate}
\item P.J. Olver, Applications of Lie Groups to Differential Equations
    (Springer-Verlag, New York, 1986); P.Winternitz, in: Partially
    Integrable Evolution Equations in Physics, eds. R.Conte and
    N.Boccara, NATO ASI Ser. C, v.310, p.515 (Kluwer Academic Publ.,
    Dordrecht, 1990).
\item P.Forg\'{a}cs and N.S.Manton, Commun.Math.Phys. 72 (1980) 15;
      J.Harnad, S.Snider and L.Vinet, J.Math.Phys. 21 (1980) 2719;
      R.Jackiw and N.S.Manton, Ann.Phys. 127 (1980) 257.
\item M.Legar\'{e} and A.D.Popov, Phys.Lett. A 198 (1995) 195;
      JETP Lett. 59 (1994) 883;\\
      T.A.Ivanova and A.D.Popov, Teor.Mat.Fiz. 102 (1995) 384;
      JETP Lett. 61 (1995) 150.
\item R.S.Ward, Phil.Trans.R.Soc.Lond. A315 (1985) 451;
       Lect.Notes in Phys.  280 (1987) 106;
      R.S.Ward,  in: Twistors in Mathematics and Physics, eds.
      T.N. Bailey and R.J. Baston, London Math. Society Lect. Note
      Ser., v. 156, p.246 (Cambridge University Press, Cambridge, 1990).
\item L.J.Mason and G.A.J.Sparling,  Phys.Lett.A 137 (1989) 29;
      J.Geom.and Phys. 8 (1992) 243.
\item I.Bakas and D.A.Depireux, Mod.Phys.Lett.A 6 (1991) 399; 1561.
\item M.J.Ablowitz, S.Chakravarty and P.A.Clarkson, Phys.Rev.Lett. 65
      (1990)
     1085; S.Chakravarty and M.J.Ablowitz,  in: Painlev\'e Transcendents,
     their Asymptotics and Physical Applications, eds. D.Levi and
     P.Winternitz, NATO ASI Ser. B, v. 278, p.331 (Plenum Press, New York,
      1992).
\item S.Chakravarty , S.Kent and E.T.Newman, J.Math.Phys. 31 (1990) 2253;
      33 (1992) 382; I.A.B.Stra\-chan,  Phys.Lett.A 154 (1991) 123;
      T.A.Ivanova and A.D.Popov, Lett. Math. Phys. 23 (1991) 29.
\item T.A.Ivanova and A.D.Popov, Phys.Lett.A 170 (1992) 293.
\item  R.S.Ward, Gen.Rel.Grav. 15 (1983) 105; N.M.J.Woodhouse, Class.
   Quantum Grav. 4 (1987) 799; 6 (1989) 933;
      N.M.J.Woodhouse and L.J.Mason, Nonlinearity 1 (1988) 73;
     J.Fletcher and N.M.J.Woodhouse,  in: Twistors in Mathematics and
      Physics, eds. T.N. Bailey, R.J. Baston, London Math. Society
     Lect. Note Ser., v. 156, p. 260  (Cambridge University
     Press, Cambridge, 1990).
\item L.J.Mason and N.M.J.Woodhouse,  Nonlinearity 6 (1993) 569.
\item J.Tafel, J.Math.Phys. 34 (1993) 1892.
\item M.Kovalyov, M.Legar\'e and L.Gagnon,  J.Math.Phys. 34 (1993) 3245.
\item M.J.Ablowitz and P.A.Clarkson, Solitons, Nonlinear Evolution
      Equations and Inverse Scattering (Cambrige University Press,
      Cambridge, 1991).
\item M.F.Atiyah, Classical Geometry of Yang-Mills Fields
      (Scuola Normale Superiore, Pisa, 1979);
      M.Atiyah and N.Hitchin, The Geometry and Dynamics of Magnetic
      Monopoles (Princeton University Press, Princeton, 1988);
      R.S.Ward and R.O.Wells Jr., Twistor Geometry and Field Theory
       (Cambridge University Press, Cambridge, 1990).
\item R.S.Ward, J.Math.Phys. 29 (1988) 386; Nonlinearity 1 (1988) 671;
      J.Math.Phys. 30 (1989) 2246; Commun.Math.Phys. 128 (1990) 319.
\item H.Ooguri and  C.Vafa, Mod.Phys.Lett. A5 (1990) 1389;  Nucl. Phys.
      B361 (1991) 469; B367 (1991) 83; W.Siegel, Phys.Rev.Lett.
      69 (1992) 1493; Phys.Rev. D46 (1992) 3235; D47 (1993) 2504; D47
      (1993) 2512.
\item A.Parkes, Nucl.Phys. B376 (1992) 279;
      S.V.Ketov, H.Nishino and S.J.Gates Jr., Phys. Lett. B307 (1993) 331;
      Nucl.Phys. B393 (1993) 149.
\item N.Berkovits and C.Vafa, Mod.Phys.Lett. A9 (1994) 653;
      Nucl.Phys. B433 (1995) 123.
\item N.Berkovits, Super-Poincar\'{e} invariant superstring field theory,
      hep-th/9503009; New spacetime-supersymmetric methods for the
      superstring, hep-th/9506036.
\item A.A.Belavin and V.E.Zakharov, Phys.Lett. B73 (1978) 53.
\item R.S.Ward, Phys.Lett. A61 (1977) 81.
\item A.Lichnerowicz,  G\'eom\'etrie des groupes de transformations (Dunod,
    Paris, 1958); S.Kobayashi,  Transformation groups in differential
    geometry (Springer-Verlag, Berlin, 1972).
\item S.V.Manakov and V.E.Zakharov, Lett.Math.Phys. 5 (1981) 247.
\item V.A.Belinsky and V.E.Zakharov, Zh.Eksp.Teor.Fiz. 75 (1978) 1953;
    77 (1979) 3.
\item S.P.Burtsev, V.E.Zakharov and A.V.Mikhailov, Teor.Mat.Fiz. 70 (1987)
      323.
\item A.P.Fordy and P.P.Kulish, Commun.Math.Phys. 89 (1983) 427.
\item F.Calogero and A.Degasperis, Commun.Math.Phys. 63 (1978) 155.
\item A.Fordy, S.Wojciechowski and I.Marshall, Phys.Lett. 113A (1986) 395.
\item A.T.Fomenko and V.V.Trofimov, Integrable systems on Lie algebras
     and symmetric spaces (Gordon and Breach, New York, 1988).
\item J.Gibbons and Th.Hermsen, Physica 11D (1984) 337.
\item E.Billey, J.Avan and O.Babelon, Phys.Lett. A186 (1994) 114; A188
      (1994) 263.
\item S.Wojciechowski, Phys.Lett. A111 (1985) 101.
\item A.C.Newell, Solitons in Mathematics and Physics
     (SIAM, Philadelphia, 1985).
\end{enumerate}
\end{document}